\documentclass[useAMS,usenatbib]{mn2e}

\usepackage{times}

\usepackage{epsfig}
\usepackage{subfigure}

\usepackage{amsmath}
\usepackage{amssymb}
\usepackage{graphicx}
\usepackage{latexsym}
\usepackage{aas_macros}

\usepackage{url}

\newcommand{\paren}[1]{\left ( #1 \right )}

\newcommand{\curly}[1]{\left \{ #1 \right \} }

\newcommand{\parenfrac}[2]{\paren{\frac{#1}{#2}}}

\newcommand{\pderiv}[2]{\frac{\partial {#1}}{\partial {#2}}}

\newcommand{\mvect}[1]{\mathbf{#1}}

\newcommand{\grad}[1]{\nabla #1}
\newcommand{\diverge}[1]{\grad{} \cdot {#1}}

\begin{document}

\title[An alternative approach to viscosity]{An alternative approach to viscosity in an accretion disc}
\author[R.~G.~Edgar]{Richard Edgar$^1$\thanks{rge21@pas.rochester.edu}  \\
$^1$Department of Physics and Astronomy, University of Rochester, Rochester, NY 14627} 
\date{\today}

\pagerange {\pageref{firstpage}--\pageref{lastpage}}

\label{firstpage}

% ----------------

\maketitle

\begin{abstract}
Purely hydrodynamic numerical experiments into the evolution of astrophysical discs typically include some sort of viscosity in order to cause accretion.
In this paper, we demonstrate an alternative method of implementing viscous forces, with extremely good angular momentum conservation properties.
The method is based on altering the cell fluxes, rather than incorporating a viscous force.
We test this method on the classical `ring spreading' problem, and demonstrate angular momentum conservation at the $10^{-8}$ level.
\end{abstract}

% -------------------


% -------------------

\section{Introduction}
\label{sec:intro}

Gas dynamics dominates the physics of many areas of astronomy.
Unfortunately, gas flow is a complicated subject, which has proven stubbornly resilient to analytic study.
This is due to the non-linear nature of the Navier-Stokes equations which describe the flow.
We are forced to look to computers to conduct numerical experiments, in order to better understand the cosmos.
The reason that computational work is better described as a `numerical experiment' rather than a `simulation' is the plethora of techniques available, with no obviously `right' method.
A recent comparison of different codes on is the work of \citet{2006MNRAS.370..529D}, who conducted a comparison of a variety of codes on the planet--disc interaction problem.
In this paper, we shall introduce an alternative method of implementing viscosity in a hydrodynamics code, and test it in the case of an accretion disc.

Viscous processes in astrophysics are typically thought to originate from magnetic effects \citep[e.g. the Magnetorotational Instability (MRI) of][ is thought to dominate transport in most accretion discs]{1991ApJ...376..214B}, rather than purely hydrodynamic processes (which would be far too weak, given the low densities involved).
However, magnetohydrodynamic (MHD) calculations are computationally expensive.
It is therefore common to incorporate conventional viscous terms into the equations used, to approximate MHD effects.
Observational effects of accretion can thereby be studied, without the added complexity of a full MHD calculation.
However, since the details of MHD transport will certainly be different from a simple physical viscosity, MHD calculations are required to verify results from the physical viscosity approach.
This has been demonstrated by \citet{2005A&A...443.1067N} in the context of the migration of low mass planets, and by \citet{2003ApJ...589..543W} for planets which can open a gap in a circumstellar disc.
A further reason to incorporate viscosity into a hydrodynamics code is to ensure that results are not dominated by details of the algorithm.
All codes for solving the equations of hydrodynamics involve some numerical dissipation.
This dissipation is dependent on both the algorithm used, and the resolution computed, and there is no general method of calculating what effect it has.
For this reason it is best not to refer to the dissipation as `numerical viscosity,' since the effect is unlikely to be precisely that of a diffusion equation.
By introducing some physical viscosity into the hydrodynamic equations, we can hope to overpower the unknown dissipation a parameter we can control.\footnote{References to `numerical viscosity' are better read as `minimum level of physical viscosity required to dominate intrinsic numerical dissipation'}
Of course, if too much physical viscosity is required to achieve this, then the quality of our numerical method is called into question.

The structure of this paper is as follows:
in section~\ref{sec:hydroeqns} we introduce the equations of hydrodynamics, and discuss how viscosity has been implemented previously in section~\ref{sec:prevwork}.
Our new approach to incorporating viscous terms is summarised in section~\ref{sec:newapproach}.
We discuss implementation details in section~\ref{sec:implementation} and our tests in section~\ref{sec:tests}.

\section{The Hydrodynamic Equations}
\label{sec:hydroeqns}

The equations of hydrodynamics have been known for several centuries, and may be written in a variety of forms.
A common one is the following:
\begin{eqnarray}
\pderiv{\rho}{t} + \diverge{(\rho \mvect{v})} & = & 0 
\label{eq:massconserve} \\
\pderiv{\mvect{v}}{t} + (\mvect{v} \cdot \nabla ) \mvect{v} & = & - \frac{1}{\rho}\grad{p} - \grad{\Phi} + \diverge{T_{ij}}
\label{eq:momconserve}
\end{eqnarray}
where $\rho$ is the density, $\mvect{v}$ the velocity of the fluid, $p$ the pressure, $\Phi$ the gravitational potential and $T_{ij}$ is the viscous stress tensor \citep[see, e.g.][]{LandauLifshitz,BatchelorFluids}. 
The first equation describes the conservation of mass and the second, conservation of momentum.
An equation of state closes the system of equations, and additional terms
may be added as required.
The viscous stress tensor is written as
\begin{equation}
T_{ij} = \nu \paren{\grad{} \mvect{v} + (\grad{} \mvect{v})^{T}} +
         \paren{\zeta - \frac{2}{3}\nu} \paren{\diverge{\mathbf{v}}} \mathbf{I}
\label{eq:ViscStressTensor}
\end{equation}
where $\nu$ is the kinematic shear viscosity, and $\zeta$ is the kinematic bulk viscosity.\footnote{If Equation~\ref{eq:momconserve} is written in terms of momentum, then these coefficients must be multiplied by the density, and $\eta \equiv \rho \nu$ is sometimes used}
The first term of Equation~\ref{eq:ViscStressTensor} represents the resistance of the fluid to shear forces, while the second measures the resistance to dilation.
Note that the first term is symmetric, which ensures that solid body rotation does not give rise to viscous forces.

\section{Previous Work}
\label{sec:prevwork}

Computer codes designed to study astrophysical fluids have included viscosity for a number of years.
We shall review the methods used in this section, starting with a brief review of how such codes work.
This discussion concentrates on time-explicit grid-based codes.
Other types of code exist, notably Smoothed Particle Hydrodynamics \cite[see, e.g.][for a description]{1990nmns.work..269B,1992ARA&A..30..543M}, but the new method for computing viscosity we are introducing here is not appropriate for them.
The following discussion is far from comprehensive (to be so would require several books), but serves as a brief outline of the principles involved.

As their name implies, grid-based codes impose a grid on the computational volume, breaking it down into cells.
The various flow variables (density, velocity, etc.) are stored for each cell.
Simulation time is advanced in a two step process.
In the \emph{source} or \emph{flux} step, the fluxes of mass, momentum etc. through each cell face are computed.
These fluxes are then used to update the cell quantities during the \emph{transport} step.
This approach enforces conservation, since the flux out of one cell will be the flux in to another.

Codes are further classified according to how they obtain their fluxes.
Some do this by direct differencing of the equations of hydrodynamics, and are often referred to as ``\textsc{Zeus}-like'' -- a reference to the \textsc{Zeus} code of \citet{1992ApJS...80..753S}.\footnote{Note that this does not imply that all these codes are derived from \textsc{Zeus}; simply that \textsc{Zeus} contains the most widely known implementation of the principles involved}
The alternative method in common use is due to Godunov, which solves a 1-D shock tube problem at every cell interface (the so-called Riemann problem).
Godunov's scheme is most commonly implemented using the Piecewise Parabolic Method (PPM) of \cite{1984JCoPh..54..174C}.
All explicit schemes for numerical hydrodynamics are subject to the Courant-Friedrichs-Lewy (CFL) condition for their maximum timestep.
This requires that
\begin{equation}
\Delta t < \frac{\Delta x}{|v_x|+c_s}
\end{equation}
Physically, this means that information must not propagate more than one grid cell per timestep.

In codes like these, viscosity has generally been implemented in one of two ways.
The first is to incorporate the viscous forces directly into the computation of the fluxes.
The acceleration of the fluid is calculated as
\begin{equation}
 - \frac{1}{\rho}\grad{p} - \grad{\Phi} + \diverge{T_{ij}}
\label{eq:viscForceInclude}
\end{equation}
where the viscosity has been added as an extra term to the usual forces of gravity and pressure gradient.
An alternative approach adds a separate substep to the source and transport steps outlined above.
In this, the velocities are evolved separately, according to
\begin{equation}
\pderiv{\mvect{v}}{t} = \diverge{T_{ij}}
\label{eq:viscSubStep}
\end{equation}
This method is used in the \textsc{Fargo} code \citep[written by][]{2000A&AS..141..165M}, and also in \textsc{Zeus} by \citet{1999MNRAS.310.1002S}.
In both cases, the viscous terms are being treated as a force.
When used explicitly, both of these methods lead to an extra CFL condition of the form
\begin{equation}
\Delta t < \frac{(\Delta x)^2}{2 \nu}
\label{eq:ViscousCFL}
\end{equation}
which must also be satisfied everywhere on the grid.
Since this can become somewhat restrictive, \citet{1989A&A...208...98K,1999MNRAS.303..696K} details an implicit solution method, which is not subject to Equation~\ref{eq:ViscousCFL}.

\section{A New Approach}
\label{sec:newapproach}

In this section, we shall discuss a new approach to incorporating viscosity into hydrodynamics codes.

Consider Equation~\ref{eq:momconserve}.
We can recast this in conservative form as
\begin{equation}
\pderiv{\rho \mvect{v}}{t} + \diverge{\rho \mvect{v} \mvect{v}} = -\grad{p} - \grad{\Phi} + \diverge{T_{ij}}
\label{eq:conserveform}
\end{equation}
The $\rho \mvect{v} \mvect{v}$ term is the momentum flux, which is supplied by the source step.
Since the divergence is taken of both this and the viscous stress tensor, an alternative formulation is suggested:
take the momentum fluxes, and subtract the viscous stress tensor.
Equation~\ref{eq:conserveform} then becomes
\begin{equation}
\pderiv{\rho \mvect{v}}{t} + \diverge{\paren{\rho \mvect{v} \mvect{v} - T_{ij}}} = -\grad{p} - \grad{\Phi}
\end{equation}
Instead of including viscosity into the source step, we should use it to generate an extra momentum flux, given by the viscous stress tensor itself.
The transport step of the code will then act on the hydrodynamic and viscous fluxes.
This has the advantage of being conservative, since whatever is added to one cell will be removed from another.
We shall see how this leads to excellent conservation of angular momentum.
However, since this method is still time-explicit, it remains subject to Equation~\ref{eq:ViscousCFL}.
As an additional bonus, this approach only requires numerical first derivatives, rather than the two derivatives required by the approach of Equations~\ref{eq:viscForceInclude} and~\ref{eq:viscSubStep}.

\section{Implementation}
\label{sec:implementation}

In this section, we shall discuss the implementation of the approach outlined in section~\ref{sec:newapproach} in a real code.
The code we shall use the \textsc{Flash} code of \citet{2000ApJS..131..273F}, an adaptive mesh refinement (AMR) code based around a PPM hydrodynamics solver.\footnote{The source code is available at \url{http://flash.uchicago.edu/}}

The downloadable source code contains a partial implementation of the method outlined in section~\ref{sec:newapproach} (indeed, it was this which inspired the present work).
However, this portion of the code is not fully implemented even in cartesians.
When performing numerical experiments into the dynamics of accretion discs, polar geometry is highly desirable.
This is because hydrodynamics codes generally conserve quantities along the grid axes.
Hence, to conserve angular momentum (rather desirable for an accretion disc), a polar grid is (almost) always required.
The downloadable source doesn't conserve angular momentum, and some extra modifications (not detailed here) were necessary to fix this problem.
We take this angular momentum conserving code as the basis of our modifications.

To compute the necessary fluxes, we must evaluate Equation~\ref{eq:ViscStressTensor} in polar co-ordinates.
This an unpleasant exercise, but is built on standard results.
We find that the six independent components of $T_{ij}$ are:
\begin{eqnarray}
T_{rr} & = & 2 \nu \pderiv{v_r}{r} + \paren{\zeta - \frac{2}{3}\nu} \paren{\diverge{\mathbf{v}}} \label{eq:Trr} \\
T_{r\phi} \equiv T_{\phi r} & = & \nu \paren{\pderiv{v_{\phi}}{r} + \frac{1}{r}\pderiv{v_r}{\phi} - \frac{v_{\phi}}{r}} \label{eq:Trphi} \\
T_{rz} \equiv T_{zr} & = & \nu \paren{\pderiv{v_z}{r} + \pderiv{v_r}{z}} \label {eq:Trz} \\
T_{\phi \phi} & = & 2 \nu \paren{\frac{1}{r}\pderiv{v_{\phi}}{\phi} + \frac{v_r}{r}} + \paren{\zeta - \frac{2}{3}\nu} \paren{\diverge{\mathbf{v}}} \label{eq:Tphiphi} \\
T_{\phi z} \equiv T_{z \phi} & = & \nu \paren{\frac{1}{r}\pderiv{v_z}{\phi} + \pderiv{v_{\phi}}{z}} \label{eq:Tphiz} \\
T_{zz} & = & 2 \nu \pderiv{v_z}{z} + \paren{\zeta - \frac{2}{3}\nu} \paren{\diverge{\mathbf{v}}} \label{eq:Tzz}
\end{eqnarray}
where
\begin{equation}
\diverge{\mvect{v}} = \frac{1}{r}\pderiv{r v_r}{r} + \frac{1}{r}\pderiv{v_{\phi}}{\phi} + \pderiv{v_z}{z}
\end{equation}
From this point on, we shall concentrate on the 2D non-axisymmetric $(r,\phi)$ case.
We therefore take $z = v_z = \partial_z = 0$, which results in a considerable simplification of these equations.

\begin{figure}
\begin{center}
\includegraphics[scale=0.5]{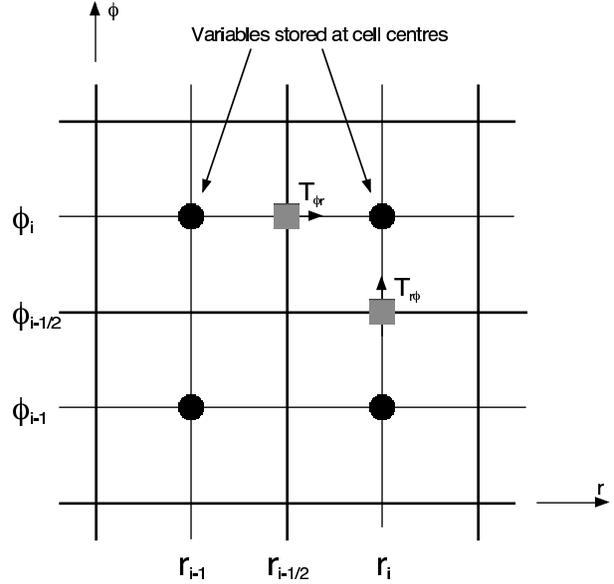}
\end{center}
\caption{Storage of variables in \textsc{Flash}.
All variables are stored at cell centres, located at $\{ (r_{i}, \phi_{j}), ( r_{i+1}, \phi_{j} ) \ldots \}$.
The cell faces are located at the `half' positions.
We require the components of the stress tensor to be centred on the cell faces}
\label{fig:FlashStorage}
\end{figure}

At this point, numerical details intrude.
In \textsc{Flash}, all flow quantities (density, velocity etc.) are stored at cell centres.
We wish to compute fluxes on the cell faces, and this introduces complications.
In particular, although $T_{r\phi} = T_{\phi r}$ mathematically, we wish to evaluate them in different places.
We require $T_{r\phi}$ (that is, the flux of $v_r$ in the $\phi$ direction) to be evaluated on the centre of the $\phi$ faces, while $T_{\phi r}$ (the flux of $v_{\phi}$ in the $r$ direction) is required on the centre of the $r$ faces.
This is illustrated in Figure~\ref{fig:FlashStorage}.

We use centred derivatives, in order to achieve second order accuracy.
For certain terms, this is easy.
Consider the first term of $T_{\phi r}$:
\begin{equation}
\left . \pderiv{v_{\phi}}{r} \right |_{i-\frac{1}{2}}
\approx
\frac{v_{\phi}^{i,j} - v_{\phi}^{i-1,j}}{r_i - r_{i-1}}
\end{equation}
The second term is more troublesome.
We want the derivative in the $\phi$ direction, but centred on an $r$ face.
We solve this problem by evaluating the derivative at the cell centres on either side of the interface, and then taking the mean:
\begin{equation}
\left . \frac{1}{r}\pderiv{v_r}{\phi} \right |_{i-\frac{1}{2}}
\approx
\frac{Q_1 + Q_2}{2} \cdot \frac{2}{r_{i-1}+r_i}
\label{eq:derivativecentring}
\end{equation}
where
\begin{eqnarray}
Q_1 & = & \frac{v_r^{i-1,j+1}-v_r^{i-1,j-1}}{\phi_{j+1} - \phi_{j-1}} \\
Q_2 & = & \frac{v_r^{i,j+1}-v_r^{i,j-1}}{\phi_{j+1} - \phi_{j-1}}
\end{eqnarray}
The final term is comparatively straightforward:
\begin{equation}
\left . \frac{v_{\phi}}{r} \right |_{i-\frac{1}{2}}
\approx
\frac{v_{\phi}^{i-1,j}+v_{\phi}^{i,j}}{r_{i-1}+r_i}
\end{equation}
where the factors of two in the averages cancel top and bottom.
Similar considerations apply when evaluating $T_{r \phi}$, although it is the first term which is trickier to centre correctly.
Note that the differencing formul\ae{} given here break down close to $r=0$ (formally, once $\Delta r \approx r$).
It would be possible to remedy this, but we shall not worry about the problem here.

Since it is actually momentum density which is transported, all the computed fluxes must be multiplied by the density at the interface.
This is computed as the mean of the density in the cells sharing the interface.
Conversion to angular momentum is handled elsewhere in the code, and so does not affect the expressions given here.
Boundary conditions are quite simple: no extra flux is added to faces on the edges of the computational volume.

\section{Tests}
\label{sec:tests}

In this section, we shall discuss some tests of the new approach to implementing viscosity discussed above.

% -------------------
\subsection{The Viscously Spreading Ring}

The canonical test of viscosity in a computer code is arguably the viscously spreading ring described by \citet{1981ARA&A..19..137P}.
In this test, the surface density evolution of material in a constant viscosity Keplerian accretion disc is monitored.
If the surface density, $\Sigma$, is initially a $\delta$-function at $r=r_0$, then at later times it is described by
\begin{equation}
\Sigma(x,\tau)
\propto
\frac{1}{\tau x^{\frac{1}{4}}}
\cdot
\exp \curly{-\frac{1+x^2}{\tau}}
\cdot
I_{\frac{1}{4}} \parenfrac{2 x}{\tau}
\label{eq:viscringanalytic}
\end{equation}
where $x = r / r_0$, $\tau = 12 \nu t r_0^2$ and $I_{\frac{1}{4}}$ is a modified Bessel function.
The constant of proportionality is set by the mass of the ring.
For further details, see \citet{1981ARA&A..19..137P}.
Note that this only tests the $T_{r\phi}$ and $T_{\phi r}$ components of the viscous stress tensor.
Since $\delta$-functions involve infinities, which are generally accepted to be numerically difficult, we start our tests from $\tau > 0$, and follow the subsequent evolution.

To implement the spreading ring in \textsc{Flash}, we use a gas with $\gamma=1.01$ (since the Rieman solver cannot cope with an isothermal equation of state).
We reset the internal energy of the gas every timestep to maintain an aspect ratio $h/r = 0.01$, which is small enough to follow the analytic solution without introducing numerical instabilities.
We use a grid with $n_r = n_{\phi} = 128$ evenly spaced grid cells, with $0.5 < x \equiv r/r_0 < 2$ for the radial range.
The inner and outer radial boundaries are reflecting, to enforce conservation.
We initialise the surface density according to Equation~\ref{eq:viscringanalytic} with $\tau = 0.016$.

\begin{figure}
\begin{center}
\includegraphics[scale=0.6]{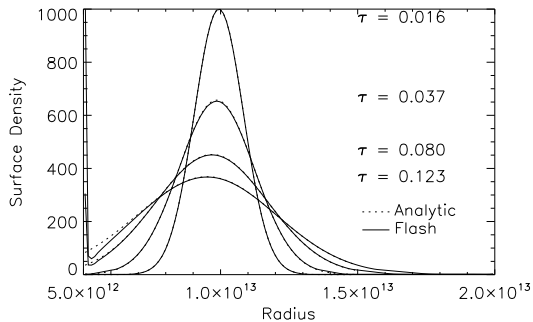}
\end{center}
\caption{Comparison of viscous ring evolution to analytic solution. Solid lines are results from \textsc{Flash}, while dotted lines show the analytic solution. The $\tau$ values corresponding to each pair of lines is also marked}
\label{fig:viscringevolve}
\end{figure}

\begin{figure}
\begin{center}
\subfigure[Mass conservation as a function of $\tau$]{\includegraphics[scale=0.6]{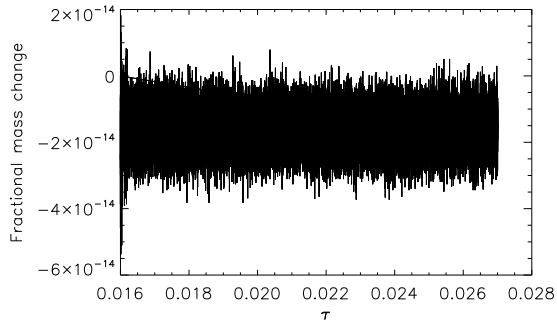}}
\subfigure[Angular momentum conservation as a function of $\tau$]{\includegraphics[scale=0.6]{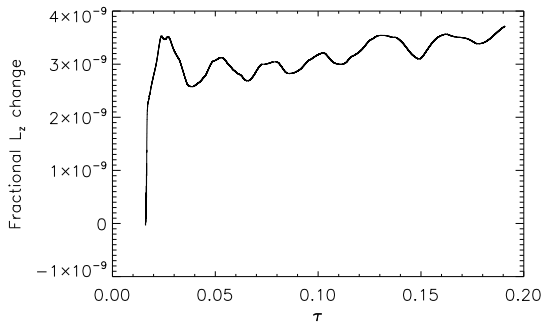}}
\end{center}
\caption{Conservation of mass and angular momentum in the \textsc{Flash} calculation show in Figure~\ref{fig:viscringevolve}}
\label{fig:viscringconserve}
\end{figure}

In Figure~\ref{fig:viscringevolve}, we compare the evolution of the ring to the theoretical solution.
Although there are some deviations, particularly once the ring encounters the inner boundary (which was reflecting, not open), the agreement is excellent.
Figure~\ref{fig:viscringconserve} demonstrates the conservation properties of the new method.
Mass is conserved to the $10^{-13}$ level (that is, machine precision), as one would expect.
The rapid oscillations are obviously rounding error from one step to the next.
Conservation of angular momentum is less good, but is still at the $10^{-8}$ level.
Computation of the viscous fluxes took approximately 2.5\% of the total CPU time.
In the absence of viscosity, angular momentum is conserved to the $10^{-13}$ level, and the ring does not spread.

From this test, we see that the new scheme is performing extremely well.

% ------------------
\subsection{Other Components}

The viscously spreading ring only tests the $T_{r\phi}$ component of the stress tensor.
Unfortunately, we do not know of any generally accepted tests of the other components.
We therefore test the other components by imposing a known velocity field, and comparing the code's computation of the stress tensor to the analytic solution.

There are three components remaining to be tested from Equations~\ref{eq:Trr} to~\ref{eq:Tzz}:
\begin{eqnarray}
S_{rr} & = & 2 \pderiv{v_r}{r} \label{eq:SrrDefine}\\
S_{\phi \phi} & = &  2 \paren{\frac{1}{r}\pderiv{v_{\phi}}{\phi} + \frac{v_r}{r}} \label{eq:SphiphiDefine} \\
B & = & \frac{1}{r}\pderiv{r v_r}{r} + \frac{1}{r}\pderiv{v_{\phi}}{\phi} \label{eq:Bdefine}
\end{eqnarray}
The Keplerian velocity field which describes the initial conditions of the viscously spreading ring makes all of these terms vanish.
We therefore adopt a perturbed field:
\begin{eqnarray}
v_r & = & v_1 \sin( m_1 \phi ) + v_2 \sin ( k r ) \\
v_{\phi} & = & \sqrt{G M_*} r^{-\frac{1}{2}} + v_3 \sin( m_3 \phi )
\end{eqnarray}
These may be easily substituted into Equations~\ref{eq:SrrDefine} to~\ref{eq:Bdefine}, to derive appropriate analytic expressions.
Evaluation of $S_{rr}$ is straightforward, since it only involves radial terms.
Computing $S_{\phi\phi}$ requires some care, to ensure that $v_r$ is evaluated on a $\phi$ face - a simple average is sufficient.
Unfortunately, $B$ must be evaluated on two different faces (similar to the complication with $T_{r\phi}$ and $T_{\phi r}$), so two separate routines are needed.

Comparing the code output to the analytic expressions, we have found that $S_{rr}$ is accurate to $< 1\%$ everywhere except the inner boundary (where differences are to be expected).
The $S_{\phi\phi}$ component is also accurate to a similar degree, everywhere except where it is close to zero (and relative accuracy becomes meaningless).
Performance evaluating $B$ was not quite so good, although still satisfactory.
Numerical acrobatics similar to those involved in Equation~\ref{eq:derivativecentring} are required to get the derivatives centred correctly.

% ------------------
\subsection{Which components are needed?}

As noted in the introduction, one of the major goals in including a physical viscosity into hydrodynamic calculations is to approximate the effects of MHD turbulence.
Unfortunately, it is not certain how best to attain this.
If the disc is to accrete, then the $T_{r\phi}$ component of the stress tensor is certainly required.
But what of the other components?
For example, in sufficiently viscous flows, convection can be shut down \citep[cf][]{2000ApJ...539..798N} if all components of the stress tensor are considered.
This is because the convective motions are subsonic (in contrast to the supersonic Keplerian shear), and hence have very low Reynolds numbers.
The effects of including different components of the stress tensor must be considered very carefully when performing numerical experiments.

\section{Conclusions}
\label{sec:conclude}

In this paper, we have presented an alternative method for including viscosity into hydrodynamics codes.
It is based on the adjustment of velocity fluxes, immediately prior to the transport step.
This approach ensures excellent conservation of angular momentum, when used in polar co-ordinates.
We have demonstrated the usefulness of this algorithm in the \textsc{Flash} code.
The computational cost is low, as compared to the Rieman solver.
Additionally, the scheme only requires first derivatives, which is numerically desirable.
Although this general approach to solving diffusion equations is not new, we are unaware of any contemporary hydrodynamics codes which make use of it.
In this paper, we have demonstrated the potential power of the technique.

% ------------

% Bibliography
\bibliography{general}
\bibliographystyle{astron}

% ------------

\section*{Acknowledgements}

The author acknowledges support from NSF grants AST-0406799, AST-0098442, AST-0406823, and NASA grants ATP04-0000-0016 and NNG04GM12G (issued through the Origins of Solar Systems Program).
This work is supported in part by the U.S. Department of Energy under Grant No. B523820 to the Center for Astrophysical Thermonuclear Flashes at the University of Chicago.
I am most grateful to Artur Gawryszczak for help with persuading \textsc{Flash} to work in polar co-ordinates.
Jonathan Dursi provided additional help with understanding the workings of \textsc{Flash}.
I am also grateful to Eric Blackman, for several fruitful discussions on the nature of viscosity, and to Gordon Ogilvie for providing a helpful (albeit unciteable) exposition of the subject in his Part III lecture notes.
Adam Frank and Alice Quillen read an early draft, and the discussion of the new method substantially improved as a result.

\bsp

\label{lastpage}

\end{document}